\documentclass{article}
 \usepackage{amsmath}
 \usepackage{authblk}
\usepackage[pdftex]{graphicx}
\usepackage{epstopdf}

\usepackage[usenames,dvipsnames,svgnames,table]{xcolor}
\newcommand{\half}{\ensuremath{^1\!\!/\!_2}}    

\begin{document}

\title{Direct observation of size scaling and elastic interaction between nano-scale defects in collision cascades}

	\author[1,2]{X. Yi}
	\author[3]{A.E. Sand}
	\author[2]{D.R. Mason\thanks{daniel.mason@ccfe.ac.uk}}
	\author[4]{M.A. Kirk}
	\author[1,2]{S.G. Roberts}
	\author[3]{K. Nordlund}
	\author[2,1]{S.L. Dudarev}
	\affil[1]{Department of Materials, University of Oxford, Parks Road, Oxford, OX1~3PH, United Kingdom}
	\affil[2]{CCFE, Culham Centre for Fusion Energy, Abingdon, Oxfordshire OX14~3DB, United Kingdom}
	\affil[3]{Department of Physics, University of Helsinki, P. O. Box 43, FI-00014, Helsinki, Finland}
	\affil[4]{Nuclear Engineering Division, Argonne National Laboratory, Argonne, Illinois 60439, USA}

\maketitle
    
    \begin{abstract}
    Using \emph{in-situ} transmission electron microscopy, we have directly observed nano-scale defects formed in ultra-high purity tungsten by low-dose high energy self-ion irradiation at 30K. At cryogenic temperature lattice defects have reduced mobility, so these microscope observations offer a window on the initial, primary damage caused by individual collision cascade events. Electron microscope images provide direct evidence for a power-law size distribution of nano-scale defects formed in high-energy cascades, with an upper size limit independent of the incident ion energy, as predicted by Sand \emph{et al.} [{\em Eur. Phys. Lett.}, {\bf{103}}:46003, (2013)]. Furthermore, the analysis of pair distribution functions of defects observed in the micrographs shows significant intra-cascade spatial correlations consistent with strong elastic interaction between the defects.
    \end{abstract}
    
    \noindent{\it PAC 61.72.J-,61.80.Az,68.37.Lp}

%
%
\section{Introduction}
Low dose neutron irradiation and plasma exposure gives rise to major changes in mechanical and thermal properties of plasma-facing components for ITER and future generation fusion devices \cite{Bolt2002}. Tungsten, chosen as a plasma-facing material for the ITER divertor thanks to its high-temperature strength and thermal conductivity \cite{Rieth2011,Rieth2013}, is expected to be significantly affected by neutrons produced from D-T fusion. In addition to transmutations \cite{Gilbert2011,Gilbert2012}, neutrons initiate collision cascades in which nano-scale defects are produced \cite{Stoller2012}. Collision cascades provide a source that drives microstructural evolution of materials under irradiation, resulting in the formation of complex dislocation and defect networks. The response of irradiated tungsten to mechanical deformation or temperature gradients differs significantly from that of the unirradiated material.

Owing to the difficulty of producing intense fluxes of neutrons with a fusion energy spectrum, self-ion irradiation is used to simulate neutron damage, and there has been recent work investigating individual ion cascade events in tungsten both experimentally\cite{Yi_PM_2013,Mason2014} and theoretically\cite{Fikar_JNM2009,Troev2011,Sand_EPL2013,Mason2014}.
As any ion moving with over the threshold displacement energy can stochastically set in motion other ions, the structure of collision cascades is fractal in nature\cite{MorenoMarin_NIMB1990}. Furthermore, at high energies cascades branch into subcascades \cite{Simeone_PRE2010,Ryazanov_JNM2009}. This fractal nature suggests that defect statistics may be governed by a universal power law similar to other systems exhibiting self-organized critical behavior \cite{Turcotte1999}. Evidence for such power-law statistics has been found recently in molecular dynamics (MD) simulations of cascades in tungsten\cite{Sand_EPL2013}.
Experimental observations at room temperature are consistent with such primary damage predicted by MD and evolved with Object Kinetic Monte Carlo (OKMC) simulations\cite{Mason2014}.

From the primary damage state at 10ns after a cascade event to an observation seconds later, the timescale gap bridged by OKMC is wide. Even if physically justified and carefully calibrated, OKMC based on semi-empirical laws for defect interactions and mobilities cannot by itself validate the initial defect statistics. This problem is compounded in self-ion irradiation, where the evolution is affected by the escape of radiation induced defects to the surface: what is observed in \emph{in-situ} TEM at room temperature is merely the remnant of immobile defects.

There is currently no experimental way to make observations of individual damage events over the nano- to milli- second timescale, but it \emph{is} possible to observe a dramatically slowed microstructural evolution using self-ion irradiation combined with \emph{in-situ} transmission electron microscopy (TEM) at cryogenic temperature. In this work we observe defects formed in 150keV and 400keV W$^+$ ion cascades in ultra-high-purity tungsten foil at 30K.  At this temperature isolated clusters of point defects, be they vacancy or interstitial in nature, are generally considered to be immobile \cite{Amino_PML2011}.

We show that the distribution of sizes of visible defects measured experimentally and predicted by MD simulations of cascades in foils does indeed follow a power law, with 150keV tungsten ions producing defect distributions with an exponent in good agreement with the foil MD simulations. We show further that 400keV tungsten ions show a slightly lower exponent in agreement with bulk MD simulations\cite{Sand_EPL2013}, indicating a deeper penetration and reduced surface effects. We also examine the defect pair correlation function and demonstrate that the visible defects are not distributed homogeneously in space, but rather show significant correlation between their positions. An estimate for the elastic energy of interaction between nano-scale dislocation loops of such correlations shows that for loops with diameter $d>4$nm this energy typically exceeds 1 eV. Our observations also show there exists a maximum defect size, of order 700 point defects, which we attribute to sub-cascade branching.


\section{Experimental observations}
\label{experimental}

Ultra-high-purity tungsten sheets (typically W $>$ 99.996 wt\%) supplied by Plansee Gruppe were used in this study. 3 mm discs punched from the sheets were mechanically thinned to around 100 $\mu$m, annealed in vacuum at 1673 K for 20 hours and finally twin-jet electropolished for TEM in 0.5 wt\% NaOH aqueous solution. The heat treatment produced a \{100\}$<$011$>$ texture, and the average grain size exceeded 10 $\mu$m.  Grain boundaries and residual dislocations are therefore not expected to significantly influence radiation defect production or evolution.  Irradiations were performed \textit{in situ} on the IVEM-Tandem Facility at Argonne National Laboratory, with 150 keV and 400 keV $W^+$ ions at 30 K respectively, up to $\sim$ $10^{16}$ $W^+$m$^{-2}$  at around 15$^{\circ}$ off the zone axis of (001) grains.  Foils chosen for microscopy analysis were $\sim$100 nm in thickness.

\begin{figure*}
			\centering%
			\begin{minipage}{.33\textwidth}
			    \centering
        	    t=175.2s\newline
			  	\includegraphics[width=.8\linewidth]{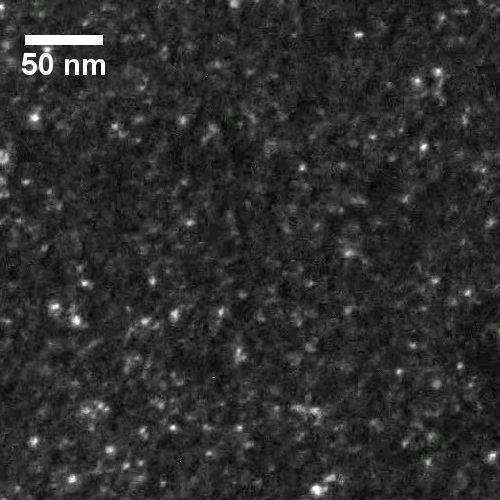}
			\end{minipage}%
			\begin{minipage}{.33\textwidth}
        	    \centering
        	    t=175.4s\newline
			  	\includegraphics[width=.8\linewidth]{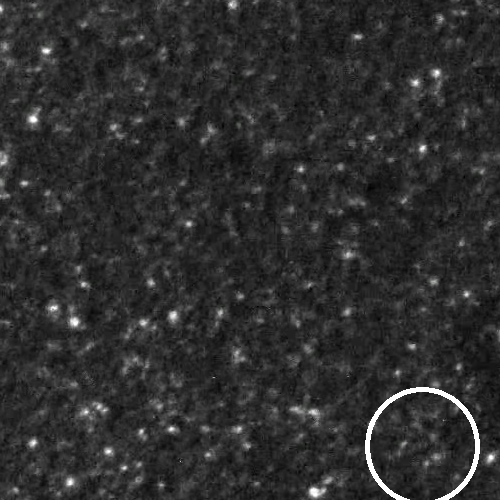}
			\end{minipage}%
			\begin{minipage}{.33\textwidth}
        	    \centering
        	    t=175.5s\newline
			  	\includegraphics[width=.8\linewidth]{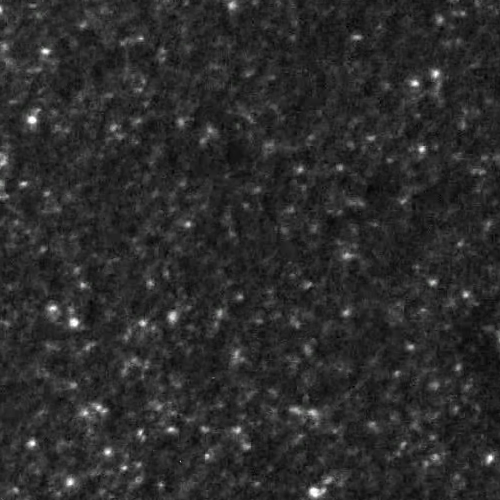}
			\end{minipage}\newline%
			\begin{minipage}{.33\textwidth}
        	    \centering
        	    t=175.7s\newline
			  	\includegraphics[width=.8\linewidth]{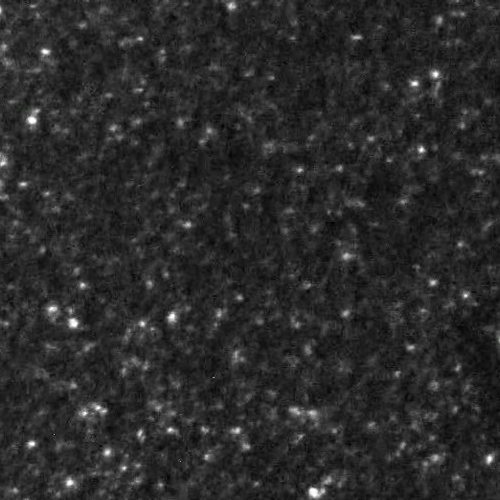}
			\end{minipage}%
			\begin{minipage}{.33\textwidth}
        	    \centering
        	    t=175.9s\newline
			  	\includegraphics[width=.8\linewidth]{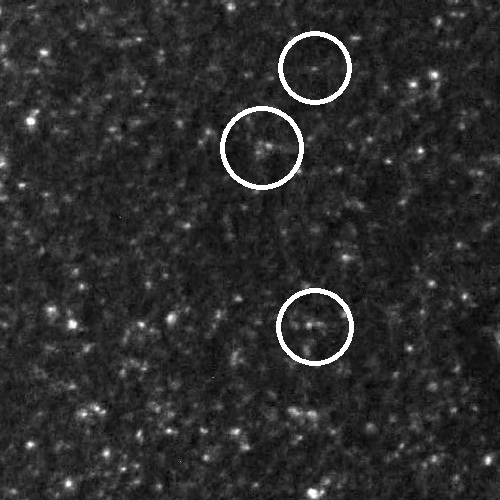}
			\end{minipage}%
			\begin{minipage}{.33\textwidth}
        	    \centering
        	    t=176.1s\newline
			  	\includegraphics[width=.8\linewidth]{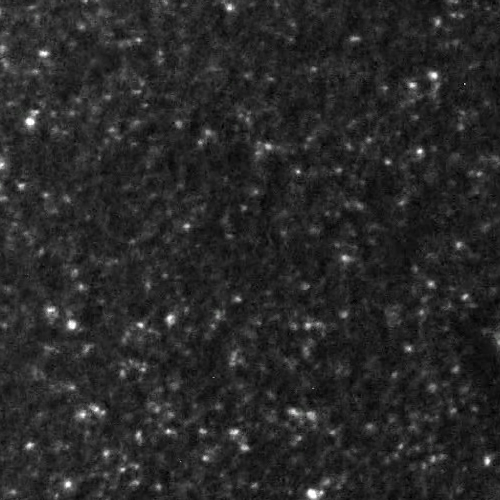}
			\end{minipage}%
           \caption{\label{fig:insitu}
            A sequence of snapshots taken over 1s irradiation with 400keV ions showing the space- and time- correlated appearance of nano-scale defects. Contrary to the case in iron, where the defects produced at low fluence are very small\cite{Stoller2012}, individual cascade events in tungsten produce correlated clusters of defects up to 9nm diameter, which appear at full brightness in a single frame and then do not subsequently evolve.
            }
		\end{figure*}

\begin{figure*}
			\centering%
			\begin{minipage}{.33\textwidth}%
			  	\includegraphics[width=.8\linewidth]{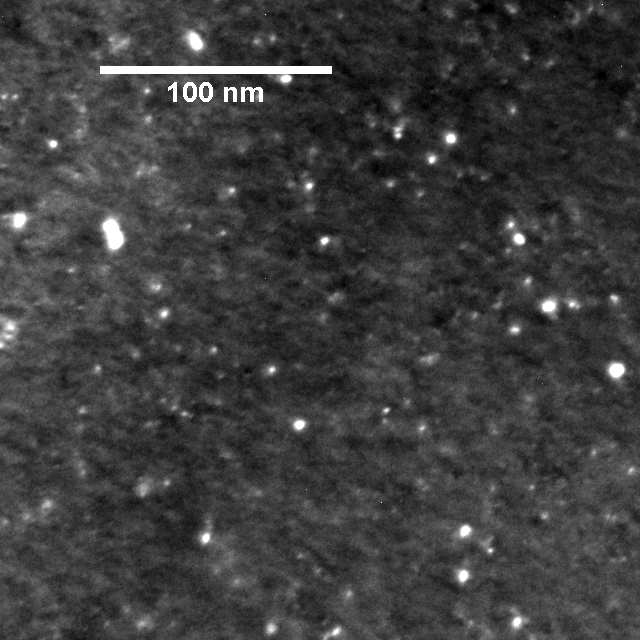}
			\end{minipage}%
			\begin{minipage}{.33\textwidth}%
			  	\includegraphics[width=.8\linewidth]{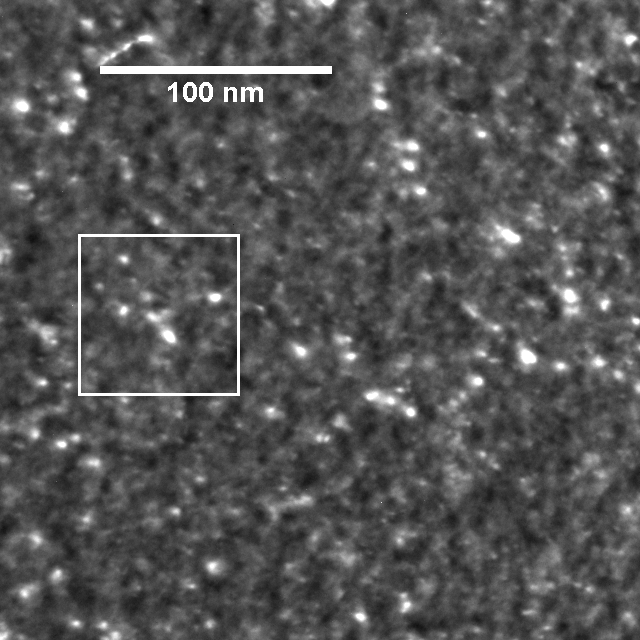}
			\end{minipage}%
			\begin{minipage}{.33\textwidth}%
			  	\includegraphics[width=.8\linewidth]{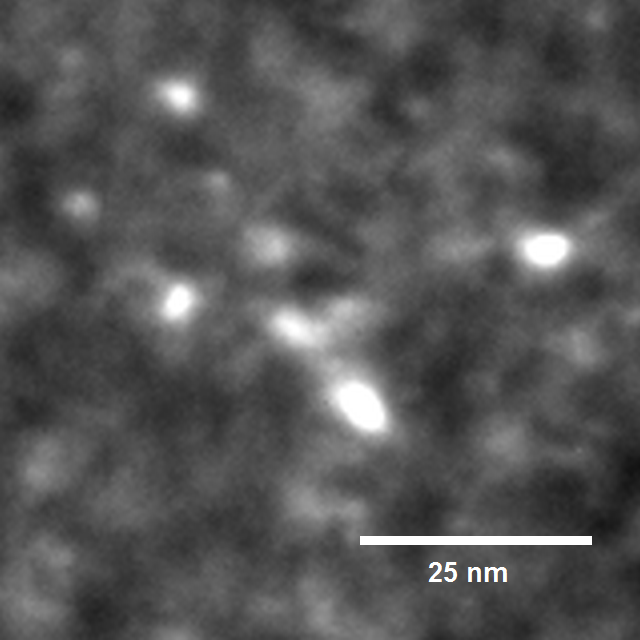}
			\end{minipage}\newline%
			\begin{minipage}{.33\textwidth}%
			  	\includegraphics[width=.8\linewidth]{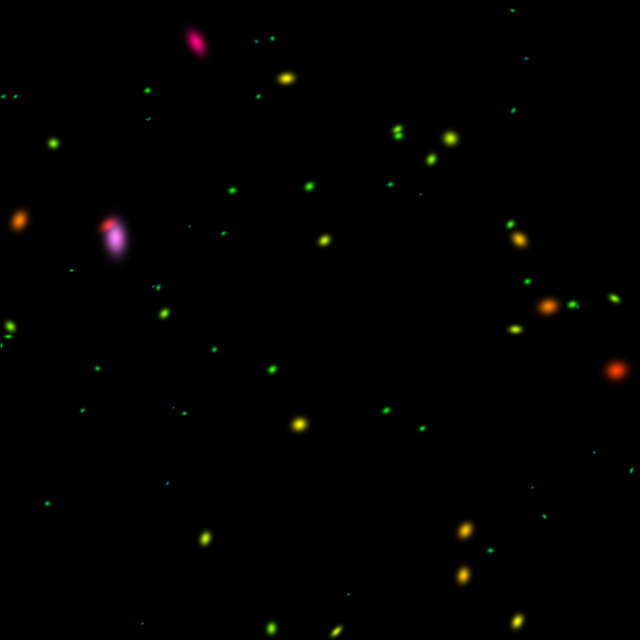}
			\end{minipage}%
			\begin{minipage}{.33\textwidth}%
			  	\includegraphics[width=.8\linewidth]{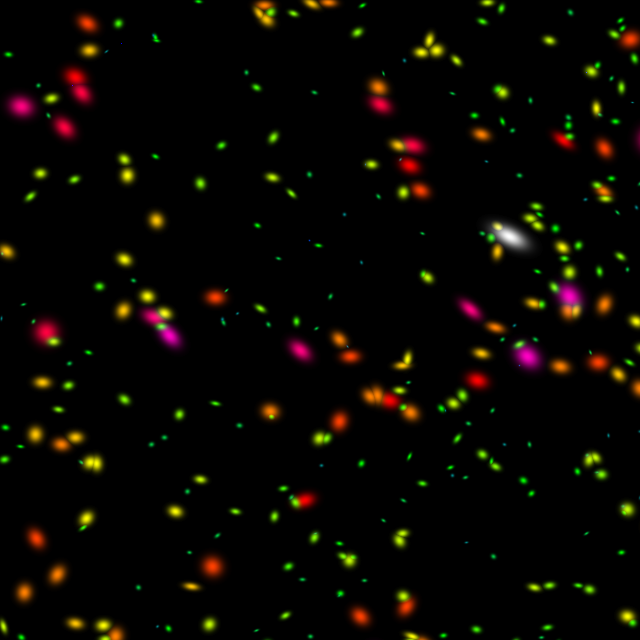}
			\end{minipage}%
			\begin{minipage}{.33\textwidth}%
			  	\includegraphics[width=.8\linewidth]{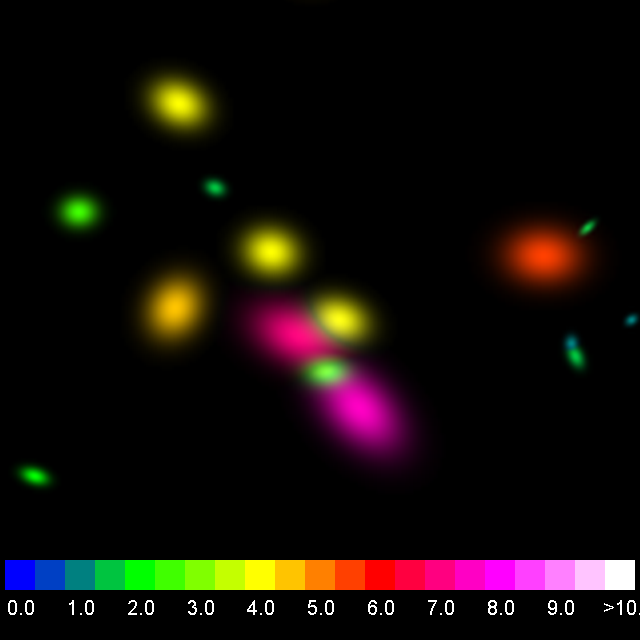}
			\end{minipage}
           \caption{\label{fig:micrograph_and_sizing} Defect clusters produced in collision cascades of tungsten. Top: Weak-beam dark-field imaging (\textbf{g} = 200, $\sim$ 3.8\textbf{g}) of defect clusters in (left, centre) 150 keV, 400 keV W$^+$ ion irradiated samples, up to a fluence of $\sim$10$^{16}$W$^+$m$^{-2}$ at 30 K. Top right: an enlarged view showing spatially correlated defects.
            Bottom: computed generated interpretations of the position and size of the defects in the images above. Defects are coloured according to their maximum diameter, with the scale (in nm) given in the coloured bar.             
            }
		\end{figure*}

Damage profiles calculated using SRIM2013 \cite{SRIM2013} monolayer damage calculation with $E_d=55.3$eV\cite{Mason2014} suggest that the peak damage for 150keV ions is 0.014dpa at 10nm depth; for 400keV ions it is 0.018dpa at 16nm depth. 
This is below the fluence for cascade overlap\cite{Yi_PM_2013}, so features we see in the micrographs can be attributed to individual cascade events.
Both ion energies are expected to produce cascades contained within the foil. Weak-beam dark-field conditions (\textbf{g} = 200, $\sim$ 3.8\textbf{g}) were selected for imaging defects. Image simulations \cite{Zhou2006} show that size of an image reflects the true defect size under imaging conditions characterized by a high excitation error, close to s = 0.2 nm$^{-1}$.
At this defect size and with these imaging conditions, loops appear as single spots and do not show a `coffee-bean' double-lobe image. Figures \ref{fig:insitu} and \ref{fig:micrograph_and_sizing} show micrographs captured \emph{in situ} and the final damage state.

\section{Simulation methods}
\label{simulation}

Molecular dynamics simulations of collision cascades resulting from self-ion irradiation in W foils were performed with the PARCAS code \cite{Nordlund1995,Nordlund1998,Ghaly1999}. Tungsten atoms with 150 keV kinetic energy were initiated above a (014) surface, 15$^{\circ}$ off the surface normal to replicate the experimental geometry and to avoid strong channeling directions, with the foil atoms initially at 0K. A simulation box of 48$\times$48$\times$65 nm was used, with periodic borders in the transverse directions and open boundaries at top and bottom surfaces. Atoms along the periodic borders were thermostated to 0 K using a Berendsen thermostat \cite{Berendsen}. A modified EAM potential \cite{Derlet_PRB2007} was used, with the repulsive part given by the universal ZBL potential \cite{SRIM2013,Bjorkas_NIMB2007}. Electronic stopping was included for atoms with kinetic energies above 10 eV \cite{Sand_EPL2013,San14}. Counting and identification of defects was performed by an automated Wigner-Seitz method \cite{Sand_EPL2013}. The simulation time (50ps) was sufficient for the final defects to be internally stable, and would require thermal migration for further recombination to occur.

\section{Results}
\label{results}

\begin{figure}
	   \includegraphics[width=55mm]{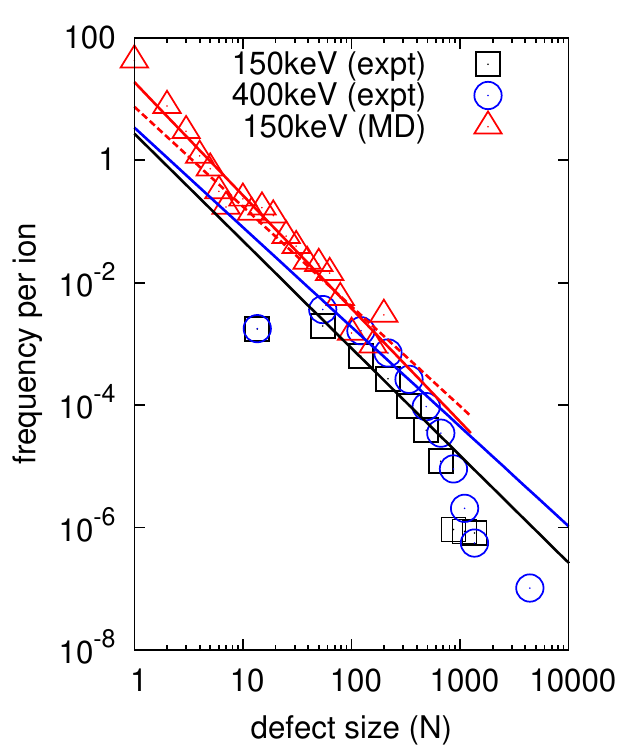}
	   \caption{\label{fig:ClusteringScalingLaw} Histogram of defect size (N) for both experiments and MD simulations. Loops smaller than 1.5nm were difficult to distinguish from the background image intensity fluctuations; the smallest size experimental bin is the accumulated frequency for diameter $d<1.5$nm. The dashed line shows the cluster power law found for 150 keV cascades in bulk tungsten \cite{Sand_EPL2013}. The histograms are normalized so that the area under the curves matches the total number of defects counted.
	   	}
	\end{figure}

The size of the defects and their relative positions were extracted from experimental micrographs using an automated procedure detailed below. For easy comparison to MD simulation results, we assume that each spot is formed by a circular $\half \langle 111 \rangle$ Burgers vector dislocation loop with diameter $d$, containing a number of point defects $N = \pi \sqrt{3} \left(  d / (2 a_0) \right)^2$, where $a_0 = 0.3165$nm is the lattice parameter of tungsten.
Measured and simulated defect size distributions were collected as histograms with varying width bins (fig. \ref{fig:ClusteringScalingLaw}).						
	
Over the loop size range $N=50-500$ point defects, corresponding to loop diameters 2-6nm in the micrographs, the observed defect size frequency from 150 keV ions is fitted to a power law $f\sim A/N^S$ with $A=2.7\pm1.4$ and $S=1.8 \pm 0.1$.
This is in good agreement with the distribution derived from 150keV foil simulations, where $A=18\pm7$ and $S=1.85 \pm 0.09$.
400keV ions produce a defect size distribution with $A=3.4\pm2.3$ and $S=1.6 \pm 0.2$. This can be compared to the frequency distribution with $A=7.5\pm1.5$ and $S=1.63 \pm 0.07$ found in MD simulations of cascades occurring in the bulk \cite{Sand_EPL2013}.

\begin{figure}
			\centering
			\begin{minipage}{.45\textwidth}
			  	\includegraphics[width=.9\linewidth]{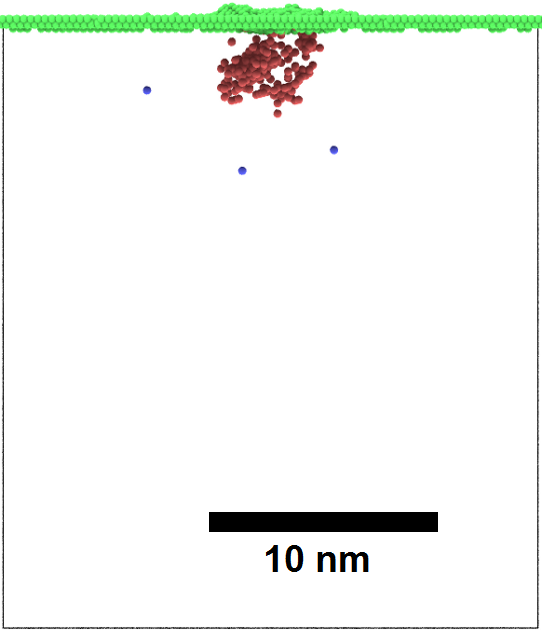}
			\end{minipage}%
			\begin{minipage}{.45\textwidth}
			  	\includegraphics[width=.9\linewidth]{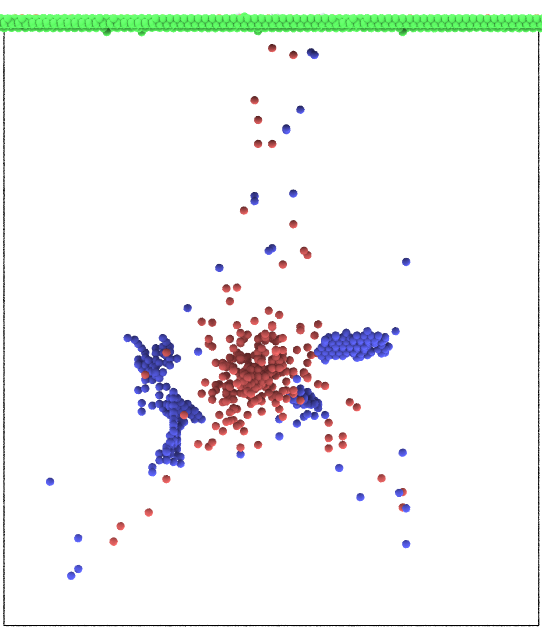}
			\end{minipage}
           \caption{\label{PrimaryDamage}
            Typical final state configurations of (left) 30keV and (right) 150keV collision cascades in tungsten foil simulated using MD. Red spheres denote vacant lattice sites, blue spheres are self-interstitials and green are surface atoms.
            }
		\end{figure}

Lower energy, 30 keV foil irradiation simulations show that most impacts result in heat spikes at very shallow depths, causing liquid-like cascade cores to erupt through the surface. This produces considerable sputtering, as well as viscous flow of the liquid atoms onto the surface \cite{Gha94}, resulting in the formation of near-surface vacancy-rich zones. Such depleted zones are seen in field ion microscope observations \cite{Wei81}, and are well reproduced by MD simulations \cite{Zho98}. Such cascades produce only a few small interstitial defects, biassing the size distribution to smaller defects. The agreement between 400 keV experimental observations and bulk MD simulations suggests that surface effects are not predominant in 400 keV ion irradiation.
Ions penetrating deeper give rise to heat spikes contained in the bulk of the material, resulting in damage which is statistically similar to bulk damage, with roughly equal numbers of interstitials and vacancies, and larger dislocation loops. Figure \ref{PrimaryDamage} shows the results of typical cascades from a 30 keV and 150 keV W ion. While only 5\% of cascades from 30 keV ions resulted in a heat spike contained beneath the surface, this proportion was 50\% for 150keV ions.

For loops containing more than $N=700$ defects, there is a deviation from linearity on the log-log plot. This deviation is significant as it indicates an upper limit to the size of defects generated in the cascades. Such a limit must exist, as it is impossible for a finite energy incident ion to generate arbitrarily large dislocation loops.

No loops larger than 10nm in diameter ($N=1300$) were observed in the 150keV irradiation experiments, and only one loop over 10nm in diameter formed during 400keV irradiation. It is not unlikely that this single very large loop was generated by the coalescence of other loops. The upper limit of the size distribution is similar for both incident ion energies, and the deviation from a power-law occurs at the same point. Such a deviation in the large event size limit is not uncommon in self-organized critical phenomena, see for example Fig. 8 of Ref. \cite{Turcotte1999}. In our case, the likely reason for the low number of very large defects is the branching of cascades into sub-cascades. This branching occurs above a certain threshold cascade energy, which for tungsten is close to 150 keV \cite{Sand_EPL2013,Ryazanov_JNM2009}.


The 150keV ion irradiations produced on average 0.27 visible loops per incident ion. This can be directly compared to 0.03 visible loops per ion in UHP W at 300K\cite{Mason2014}, indicating that at room temperature 90\% of loops are sufficiently mobile to be lost to the surface or annihilate each other. The average loop diameter at 30K is 2.6nm ($N=90$), compared to 2.0nm at room temperature, owing to the preferential annihilation of large loops\cite{Mason2014}. The 400keV self-ion irradiation experiments produced 0.55 visible loops per incident ion, with average diameter $d=2.9$nm, ($N=110$).

The number of visible (diameter $>$1.5nm) defects per cascade predicted by MD is greater than that seen in experiment; the foil simulations expected to produce 1.1 defects per ion if the power law is extended to $N=700$. On the one hand, this is not surprising given the variability associated with the choice of interatomic potentials \cite{Fikar_JNM2009, San14}. On the other hand, the short time scale of MD simulations means that even mobile defects have little time to migrate and annihilate, or reach the surface. The total number of defects found in a cascade simulation represents an upper bound to the possible number of surviving defects observed on experimental time scales. A further apparent feature of Fig. \ref{fig:ClusteringScalingLaw} is that the experimental distribution extends to significantly larger sizes than the distribution generated by MD. This unusual and interesting aspect of comparison between simulations and observations highlights the difference between sampling in the two cases. We were able to analyse only 50 MD cascades, whereas the 150keV experiment comprised 10000 cascades, and the 400keV experiment 20000 cascades. \emph{In situ} experiments therefore make it possible to observe events too infrequent to find in MD simulations. 
The average diameter of visible loops in the foil simulations is $d=2.6$nm, in agreement with experiment.
	
\begin{figure}
	   \includegraphics[width=.8\linewidth]{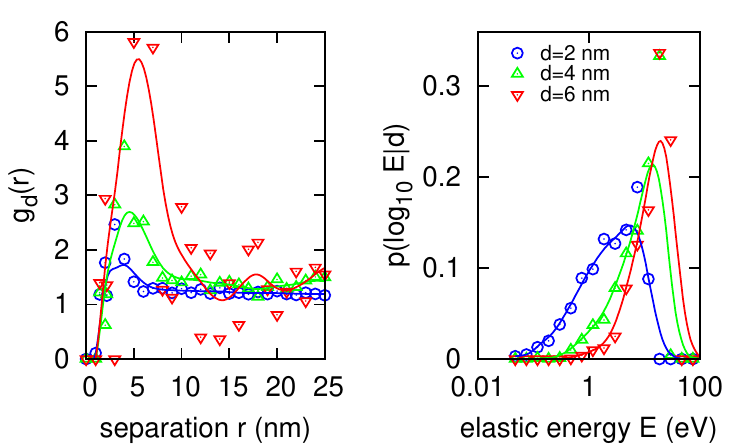}
	   \caption{\label{fig:radialDistributionFunction} Analysis of spatial correlations of defect clusters obtained from micrographs of 400 keV W$^+$ irradiated tungsten at 30 K. (a) The radial pair distribution function $g_d(r)$ is plotted against the separation distance $r$ for defects with diameters $d> 0, 2, 4, 6$ nm.  The solid lines are to guide the eye.(b) A characteristic elastic energy scale computed from the size and estimated separation of the loops (see text).}
	\end{figure}

The radial distribution function $g_d(r)$, defined in eqn.\ref{gofr} as the probability of finding pairs of spots with diameters greater than $d$, separated by distance $r$ is shown in fig. \ref{fig:radialDistributionFunction}a.		
We see a maximum correlation at $r=4$nm, for $d=2-4$nm spots in the micrographs, which is the same as the scale (4nm) of the projected radial distribution of clusters with $N>=30$ found in cascade simulations. Such large clusters occur only infrequently in MD simulations, making it difficult to draw a corresponding smooth curve for $g_{d>0}(r)$. 
The maximum of $g_{d=0}(r)$ for the MD simulations is 9nm, decreasing to 6nm if we exclude single point defects from the analysis. The average radius of gyration $r_g$, also decreases from $r_g = 10$nm for $N\ge1$ to $r_g = 4$nm for $N\ge10$. This shows that in simulations, larger defects form closer together than small defect clusters or point defects.


From the relative positions of the loops in the image we can compute a typical elastic interaction energy. While it is not possible to judge the relative depth of loops, their orientations or Burgers vectors, we can nevertheless estimate the order of magnitude. The elastic energy of interaction between two defects at distances greater than the size of the defects is
        \begin{equation}
            U_{ab}  =P^{(a)}_{ij}P^{(b)}_{kl} {\frac{\partial}{\partial R_j}}{\frac{\partial}{\partial R_l}} G_{ik}({\bf R}),
        \end{equation}
where ${\bf R}$ is the relative position of defects $a$ and $b$, $P^{(a)}_{ij}$ and $P^{(b)}_{kl}$ and the defect dipole tensors, and $G_{ik}({\bf R})$ is the elastic Green's function. Simplified expressions derived from this equation are given in Refs. \cite{Mason2014,Brown1995,Dudarev2010}.
For each loop in the 400keV irradiation images, the maximum value of this energy for each loop pair was computed.
The result (fig. \ref{fig:radialDistributionFunction}b) illustrates that the elastic energy of interaction between pairs of loops is typically greater than 1eV for loops with diameters over 4nm. This agrees with the analysis of elasticity-mediated ordering of nano-scale dislocation loops \cite{Brown1995,Dudarev2010}, and is a direct experimental confirmation of the occurrence of elastic trapping of interacting dislocation loops in cascades found in OKMC simulations\cite{Mason2014}.


\section{Conclusions}
\label{conclusion}
	In this joint experimental-theoretical work we have studied characteristics of the primary stage of radiation damage in pure tungsten foils due to self-ions.
	By performing \emph{in situ} irradiations at 30K, we have restricted the mobility of defects produced in the collision cascade.
	To a first approximation, we can claim all defects are generated in individual cascades and become immobilised rapidly after the cascade cools, and so what we see in the TEM is very close to the final stable configuration found in (an ideal) MD simulation.
	
	We have shown that defects generated in cascades can not be treated as isolated.
	Figure \ref{fig:radialDistributionFunction}a shows that the positions of dislocation loops are correlated, and fig. \ref{fig:radialDistributionFunction}b shows that this implies significant elastic forces between the loops.
	
	We have also shown that defect size distributions from our MD simulations are in very good agreement with experiment.
	In particular we have demonstrated that a power-law distribution of loop sizes holds for cascades in ion-irradiated tungsten foils as well as in bulk. The spatial distributions of simulated cascades show that large dislocation loops are generated in close proximity to each other, at distances corresponding to those seen in the TEM.


	These results show that the number of point defects produced per incident ion is not the only relevant metric for the subsequent evolution of the microstructure.
	We have shown that the distribution of loop sizes and their relative positions can be measured experimentally and are well-reproduced by MD simulation.
	This information is critical for determining whether elastic interactions are sufficient to bring defects together, forming an obstacle for further dislocation motion, or whether thermal fluctuation is sufficient for defects to find alternate sinks and be removed.

\section*{Acknowledgements}

The \textit{in situ} irradiation experiments were performed at the Nuclear Engineering Division at Argonne National Laboratory, a US Department of Energy Office of Science Laboratory operated under Contract No. DE-AC02-06CH 11357 by U Chicago Argonne, LLC. We thank P. Baldo, E. Ryan and Y. Huang for help with irradiations and microscopy. XY acknowledges funding support from Culham Center for Fusion Energy via a Research Fellowship at St. Edmund Hall, University of Oxford and from the EPSRC via the program grant `Materials for Fusion and Fission Power', EP/H018921/1. This work has been carried out within the framework of the EUROfusion Consortium and has received funding from the Euratom research and training programme 2014-2018 under grant agreement No 633053. The views and opinions expressed herein do not necessarily reflect those of the European Commission. To obtain further information on the data and models underlying this paper please contact PublicationsManager@ccfe.ac.uk.


\bibliographystyle{plain}

\begin{thebibliography}{50}


    \bibitem{Bolt2002}
    Bolt H., Barabash V., Federici G., Linke J., Loarte A., Roth J., Sato K.
     \newblock{Plasma facing and high heat flux materials  needs for ITER and beyond},
	\newblock {\em J. Nucl. Mater.}, \newblock{\bf 307-311}:43--52 (2002).

    \bibitem{Rieth2011}
    Rieth M., {\it et al.}
     \newblock{Review on the EFDA programme on tungsten materials technology and science},
	\newblock {\em J. Nucl. Mater.}, \newblock{\bf 417}:463--467 (2011).

    \bibitem{Rieth2013}
    Rieth M., {\it et al.}
     \newblock{Recent progress in research on tungsten materials for nuclear fusion applications in Europe},
	\newblock {\em J. Nucl. Mater.}, \newblock{\bf 432}:482--500 (2013).

    \bibitem{Gilbert2011}
    Gilbert, M. R., and Sublet, J.-C.,
     \newblock{Neutron-induced transmutation effects in W and W-alloys in a fusion environment},
    \newblock{\em Nucl. Fusion}, \newblock{\bf 51}:043005 (2011).
	
    \bibitem{Gilbert2012}
    Gilbert, M. R., Dudarev, S. L., Zheng, S., Packer, L. W., and Sublet, J.-C.,
     \newblock{An integrated model for materials in a fusion power plant: transmutation, gas production, and helium embrittlement under neutron irradiation},
    \newblock{\em Nucl. Fusion}, \newblock{\bf 52}:083019 (2012).

    \bibitem{Stoller2012}
    Stoller, R. E.
    \newblock{Primary radiation damage formation},
    in: \newblock{\em Comprehensive Nuclear Materials}, R. J. M. Konings (Ed.), Elsevier, 2012, pp 293--332.

   \bibitem{Yi_PM_2013}
   Yi, X., Jenkins, M.L., Briceno, M., Roberts, S.G., Zhou, Z., and Kirk, M.A.,
    \newblock {\it{In situ} study of self-ion irradiation damage in {W and W5Re} at 500C},
   \newblock {\em Phil. Mag.}, \newblock{\bf 93}:1715--1738, (2013).

    \bibitem{Mason2014}
    Mason, D.R., Yi, X.,  Kirk, M.A., and Dudarev, S.L.,
     \newblock {Elastic trapping of dislocation loops in cascades in ion-irradiated tungsten foils},
    \newblock {\em J. Phys.: Condens. Matter}, \newblock{\bf 26}:375701 (2014).

	\bibitem{Fikar_JNM2009}
	Fikar, J., and Sch\"aublin, R.,
 	\newblock{Molecular dynamics simulation of radiation damage in {BCC} tungsten},
	\newblock{\em J. Nucl. Mater.},\newblock{\bf 386--288}:97--101 (2009).

    \bibitem{Troev2011}
    Troev, T., Nankov N., and Yoshiie T.,
     \newblock {Simulation of displacement cascades in tungsten irradiated by fusion neutrons},
    \newblock {\em Nucl. Instrum. Meth. B}, \newblock{\bf{269}}:566-571, (2011).

    \bibitem{Sand_EPL2013}
    Sand, A.E., Dudarev, S.L., and Nordlund, K.,
     \newblock {High-energy collision cascades in tungsten: Dislocation loops structure and clustering scaling laws},
    \newblock {\em Eur. Phys. Lett.}, \newblock{\bf{103}}:46003, (2013).

    \bibitem{MorenoMarin_NIMB1990}
    Moreno-Marin, J.C., Conrad, U., Urbassek, H.M., and Gras-Marti, A.,
     \newblock {Fractal structure of collision cascades},
     \newblock{\em Nucl. Instrum. Meth. B},\newblock{\bf{48}}:404--407, (1990).

    \bibitem{Simeone_PRE2010}
    Simeone, D., Luneville, L., and Serruys, Y.,
     \newblock {Cascade fragmentation under ion beam irradiation: A fractal approach},
    \newblock {\em Phys. Rev. E}, \newblock{\bf{82}}:011122, (2010).

    \bibitem{Ryazanov_JNM2009}
    Ryazanov, A.I., Metelkin, E.V., and Semenov, E.V.,
     \newblock {Modeling of cascade and sub-cascade formation at high PKA energies in irradiated fusion structural materials},
    \newblock {\em J. Nucl. Mater.}, \newblock{\bf{386--388}}:132--134, (2009).

   \bibitem{Turcotte1999}
    Turcotte, D.L., 
     \newblock {Self-organized criticality},
    \newblock {\em Rep. Prog. Phys.}, \newblock{\bf 62}:1377--1429 (1999).

    \bibitem{Amino_PML2011}
    Amino, T., Arakawa, K., and Mori, H.,
     \newblock {Activation energy for long-range migration of self-interstitial atoms in tungsten obtained by direct measurement of radiation-induced point-defect clusters},
    \newblock {\em Phil. Mag. Lett}, \newblock{\bf 91}:86--96  (2011).

 \bibitem{SRIM2013}
     Ziegler, J.F., Biersack, J.P., and Littmark, U.,
    \newblock{\em The Stopping and Range of Ions in Solids},
    \newblock{Pergamon}, (1985),
    \newblock {\em http://www.srim.org/ }

  \bibitem{Zhou2006}
   Z. Zhou, M. L. Jenkins, S. L. Dudarev, A. P. Sutton, and M. A. Kirk,
     \newblock {Simulations of weak-beam diffraction contrast images of dislocation loops by the many-beam Howie-Basinski equations},
    \newblock {\em Phil. Mag}, \newblock{\bf 86}:4851--4881 (2006).

   \bibitem{Nordlund1995}
    K. Nordlund,
     \newblock {Molecular dynamics simulation of ion ranges in the 1-100 keV energy range},
    \newblock {\em Comput. Mater. Sci.}, \newblock{\bf 3}:448--456 (1995).

  \bibitem{Nordlund1998}
   K. Nordlund, M. Ghaly, R. S. Averback, M. Caturla, T. Diaz de la Rubia, and J. Tarus,
     \newblock {Defect production in collision cascades in elemental semiconductors and fcc metals},
    \newblock {\em J. Phys. Rev. B}, \newblock{\bf 57}:7556--7570 (1998).

 \bibitem{Ghaly1999}
   M. Ghaly, K. Nordlund, and R. S. Averback,
     \newblock {Molecular dynamics investigations of surface damage produced by kiloelectronvolt self-bombardment of solids},
    \newblock {\em Phil. Mag. A}, \newblock{\bf 79}:795--820 (1999).

     \bibitem{Berendsen}
      H.J.C. Berendsen, J.P.M. Postma, W.F. van Gunsteren, A. DiNola, and J.R. Haak,
       \newblock {Molecular dynamics with coupling to an external bath}
      \newblock {\em J. Chem. Phys.}, \newblock{\bf 81}:3684--3690 (1984).


    \bibitem{Brown1995}
    L.M. Brown,
     \newblock{Infinitesimal loops and tweed structure},
    \newblock{\em Czech. J. Phys.}, \newblock{\bf 45}:893--906 (1995)

    \bibitem{Dudarev2010}
    S. L. Dudarev, M. R. Gilbert, K. Arakawa, H. Mori, Z. Yao, M. L. Jenkins, and P. M. Derlet,
     \newblock {Langevin model for real-time Brownian dynamics of interacting nanodefects in irradiated metals},
    \newblock {\em Phys. Rev}, \newblock{\bf B81}:224107 (2010); S.L. Dudarev, K. Arakawa, X. Yi, Z. Yao, M.L. Jenkins, M.R. Gilbert, P.M. Derlet,
     \newblock {Spatial ordering of nano-dislocation loops in ion-irradiated materials},
    \newblock {\em J. Nucl. Mater.} \newblock{\bf 455}:16--20 (2014).

    \bibitem{Derlet_PRB2007}
    Derlet, P. M., Nguyen-Manh, D., and Dudarev, S.L.,
     \newblock {Multiscale modeling of crowdion and vacancy defects in body-centered-cubic transition metals},
    \newblock {\em Phys. Rev. B}, \newblock{\bf{76}}:054107, (2007).

    \bibitem{Bjorkas_NIMB2007}
    Bj\"orkas, C., and Nordlund, K.,
    \newblock{\em Nucl. Instrum. Meth. B}, \newblock{\bf 259}:853, (2007).

    \bibitem{San14}
      Sand, A. E., Nordlund, K. and Dudarev, S. L.,
       \newblock {Radiation damage production in massive cascades initiated by fusion neutrons in tungsten}
      \newblock {\em J. Nucl. Mater.} \newblock{\bf 455}:207--211 (2014); Sand, A.E. and Nordlund, K.,
       \newblock {On the lower energy limit of electronic stopping in simulated collision cascades in Ni, Pd and Pt}
      \newblock {\em J. Nucl. Mater.} \newblock{\bf 456}:99--105 (2015).


    \bibitem{Gha94}
      Ghaly, M. and Averback, R. S.,
       \newblock {Effect of Viscous Flow on Ion Damage near Solid Surfaces}
      \newblock {\em Phys. Rev. Lett.}, \newblock{\bf 72}:364-367 (1994).
%
%

    \bibitem{Wei81}
      Wei, C.Y., Current, M.I., and Seidman, D.N.,
       \newblock{Direct observation of the primary state of damage of ion-irradiated tungsten I. Three-dimensional spacial distribution of vacancies}
      \newblock{\em Phil. Mag. A}, \newblock{\bf 44}:459--491 (1981); Current, M.I.,  Wei, C.Y., and Seidman, D.N.,
       \newblock{Single atom sputtering events Direct observation of near-surface depleted zones in ion-irradiated tungsten}
      \newblock{\em Phil. Mag. A}, \newblock{\bf 43}:103--138 (1981).

  \bibitem{Zho98}
    Zhong, Y., Nordlund, K., Ghaly, M., and Averback, R.S.,
     \newblock{Defect production in tungsten: A comparison between field ion microscopy and molecular dynamics simulations}
    \newblock {\em Phys. Rev. B}, \newblock{\bf 58}:2361--2364 (1998).


   \bibitem{RidCal1978}
   T. W. Ridler and S. Calvard,
    \newblock {Picture thresholding using an iterative selective method},
   \newblock {\em IEEE Trans. Syst. Man. Cybern}, \newblock{\bf SMC -8}:630--632 (1978).


\end{thebibliography}

 \appendix

\section{Analysis of micrographs}
\label{analysis}
	Defects visible in the micrographs were detected with a newly developed modification of Ridler and Calvard's isodata algorithm \cite{RidCal1978}, proven against an extensive catalogue of measurements made by hand.
	Once obvious artifacts such as foil edges and screw dislocations are removed from the image, the algorithm proceeds in four stages: flattening, identifying, comparing, and counting.
	
			The flattening stage removes long-wavelength (10s to 100s nm) variations in image intensity due to variations in the foil thickness.
			For each pixel $i$ in the micrograph with intensity $f_i$, the average intensity $\bar{f}_i$ in a region 50x50nm is computed, excluding bright pixels over the Ridler and Calvard threshold\cite{RidCal1978}: $f_{\mathrm{thresh}} = \half (\bar{f}_{\mathrm{fore}} + \bar{f}_{\mathrm{back}})$.
			Then the intensity of the pixel is adjusted to
				\begin{equation}
					f_i \leftarrow \left( f_i -1 \right) \left( \frac{ 1 - \bar{f}_i }{ 1 - \bar{f} } \right),
				\end{equation}
				where $\bar{f}$ is a target background intensity.
			This flattening is repeated, averaging over a smaller 10x10nm region.

			Next all pixels with $f_i \ge f_x \equiv \half( f_{\mathrm{thresh}} + \bar{f}_{\mathrm{back}})$ are identified.
			A spot is defined as a contiguous region within this subset which includes a single maximum with intensity $f_i>f_{\mathrm{thresh}}$.
			
			A path may exist between two spots along which the intensity does not drop below $f_x$.
			If this is identified, it is necessary to determine whether we indeed have two spots, or if it is merely intensity fluctuations.
			For each pixel on a straight line connecting two maxima, if the intensity drops below 70\% of the interpolated value, or below $f_x$, the maxima are treated as two spots.
			If no point on the line drops below this threshold, the two maxima are merged.
			
			With the positions and intensities of all pixels in the spot, the maximum diameter can be computed as $d=4 r_g^+$, where $r_g^+$ is the square root of the larger eigenvalue of the tensor $\mathbf{T} = \sum_i f_i (\vec{r}_i - \langle \vec{r}\rangle)\otimes(\vec{r}_i - \langle\vec{r}\rangle)/\sum_i f_i$.
					
			If this procedure is applied to a reference image of unirradiated sample, a number of spurious dim spots will be identified, corresponding to short- wavelength variations in intensity present in all TEM micrographs.
			The mean and standard deviation of spots in the reference image is stored as a function of spot size.
			A spot in an image is accepted if its intensity passes a $\chi^2$ test at the 5\% significance level.
					
			With the position $\vec{r}_a$ and diameter $d_a$ computed for each spot as above, counting is straightforward.
			We add to the analysis the pairwise distribution $g_d(r)$, defined as
				\begin{equation}
				    \label{gofr}
					g_d(r) = \frac{1}{Z} \frac{1}{2} \sum_{a,b \neq a} H(d_a - d) H(d_b - d) \delta( \left| \vec{r}_b - \vec{r}_a \right| - r ) ,
				\end{equation}
			i.e. a count of all pairs of spots which both have diameters greater than $d$ and are separated by distance $r$.
			$H(x)$ is the Heaviside function.
			The normalisation $Z$ is chosen such that the expectation value of $g_d(r)$ is one if the spots are randomly placed, computed numerically by randomly repositioning the spots a large number of times.

\end{document}